\documentclass{article}
\usepackage[T1]{fontenc}
\usepackage[utf8]{inputenc}
\IfFileExists{ismir.sty}{%
\usepackage{ismir} 
}{%
  \usepackage[a4paper,margin=2cm]{geometry}
  \usepackage{lineno}
  \newcommand{\conferenceyear}{2026}

}
\usepackage{amsmath,amssymb,booktabs,url}
\usepackage{graphicx}
\usepackage{multirow}

\title{Beyond Piano: Cross-Instrument MIDI Velocity Estimation via Differentiable SoundFont Proxies}

\multauthor
  {Zhanhong He$^1$ \hspace{1cm} Hanyu Meng$^2$ \hspace{1cm} David (Defeng) Huang$^1$ \hspace{1cm} Roberto Togneri $^1$}
  {
  $^1$ University of Western Australia, Perth, Australia\\
  $^2$ University of New South Wales, Sydney, Australia\\
  {\tt\small zhanh.he.uwa@gmail.com, hanyu.meng@unsw.edu.au, \{david.huang; roberto.togneri\}@uwa.edu.au}
  }

\def\authorname{Zhanhong He, Hanyu Meng, David (Defeng) Huang, Roberto Togneri}

\begin{document}
\maketitle
\ifdefined\linenumbers\linenumbers\fi

\begin{abstract}
Many music datasets contain MIDI notes but lack reliable velocities, defaulting to a constant value. This absence is especially problematic outside the piano domain, as velocity is a core component for expressive rendering, music generation, and performance analysis. This paper studies cross-instrument MIDI velocity estimation in this label-scarce setting. Starting from a piano-trained velocity estimator, we recast target-instrument adaptation as predicting renderer-conditioned velocities whose rendering matches the dynamics of the performance audio. This adaptation can be driven by either differentiable synthesizers (Diff-Synth) or our proposed differentiable SoundFont proxies (Diff-SFProxy). We highlight the Diff-SFProxy: it supervises velocity through note-wise, loudness-related acoustic parameters rather than waveform reconstruction, focusing gradients on velocity-dependent behavior. Experiments on piano and guitar show that Diff-SFProxy is effective for cross-instrument MIDI velocity estimation, while waveform-domain Diff-Synth degrades performance.
\end{abstract}



\section{Introduction}
MIDI velocity is the per-note control parameter in symbolic music systems, controlling expressive aspects of rendering such as amplitude and timbre. Since its effect varies across instruments and recording conditions, velocity is related to but cannot be directly replaced by loudness measurement \cite{dan2006psy,jeong2018timbre}. Reliable velocity labels are crucial for expressive rendering \cite{rhyu2022sketching, zhang2024dexter}, music generation \cite{tang2025integrated, borovik2025symupe}, performance analysis \cite{morsi2025enable_rach3} and data curation \cite{zhang2022atepp}. However, most publicly available MIDI files lack velocity corresponding to human performance. A common practice in music software is to map score-level dynamic markings (such as $p$ and $f$) to fixed velocities, or default to 64 when unavailable \cite{dan2006psy}. Such velocities carry minimal expressive information. By contrast, learning-based velocity estimation from performance audio works well, but almost entirely in the piano domain \cite{jeong2018timbre, simonetta2022acoustics, kim2024method, he2026score}. The reason is data: piano datasets such as MAESTRO \cite{hawthorne2019maestro} are enabled by sensor-equipped Yamaha Disklavier pianos, which provide note-aligned velocity labels from recorded performances. For guitar, violin, and other instruments, public datasets \cite{tamer2023violin,riley2024gaps,riley2024high} provide audio and aligned notes but no velocity labels.

This paper studies label-scarce cross-instrument MIDI velocity estimation. In music transcription, a common strategy is to train on synthesized audio, where symbolic labels are available by construction, and transfer the model to real performances \cite{maman2022unaligned, zang2024synthtab, kusaka2025learn}. Recent works have
used this strategy for synthetic velocity supervision \cite{sato2024annotationfree, wang2026vioptt}, but the synthetic-to-real gap remains. We instead take a more direct route. Starting from a velocity estimator (VeloEst) trained on piano, we adapt it to target instruments using real performance audio, predicting velocities whose rendering matches the dynamics of the recording.

\begin{figure*}[htbp]
\centering
\includegraphics[width=\textwidth]{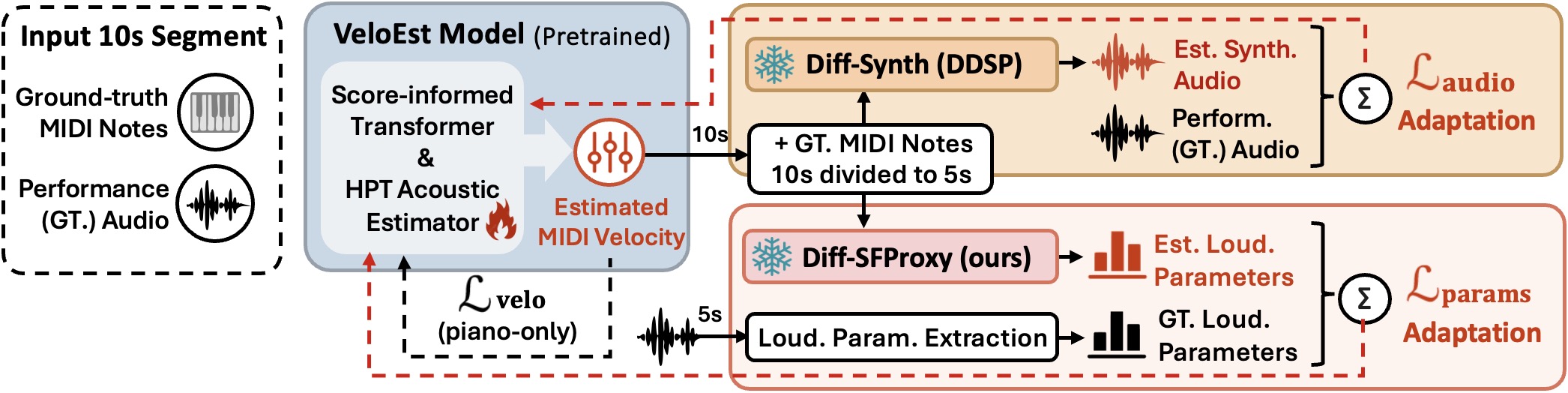}
\vspace{-18pt}
\caption{Overview of the cross-instrument velocity estimation framework. VeloEst is trained with ground-truth velocity labels on piano ($\mathcal{L}_{\mathrm{velo}}$), then fine-tuned for target instruments through one of two adaptation strategies. Diff-Synth provides waveform-domain supervision ($\mathcal{L}_{\mathrm{audio}}$). Diff-SFProxy provides note-wise loudness-parameter supervision ($\mathcal{L}_{\mathrm{params}}$).}
\label{fig1}
\end{figure*}

For Diff-Synth, while several implementations exist for each instrument, we adopt DDSP-Piano \cite{renault2023ddsp_piano} and DDSP-Guitar-Synth \cite{jonason2024guitar}, both of which are MIDI-oriented. Velocity estimates are rendered into audio with aligned note events, and an audio-based loss against the real recording drives gradients back into VeloEst. Yet Diff-Synth only approximates the real instrument, so waveform or spectral objectives spend much of their gradient on residual timbral mismatch rather than on velocity-dependent behavior.

Our proposed Diff-SFProxy follows the neural-proxy strategy for black-box audio systems \cite{martinez2021differentiable, barkan2023is2, combes2025neural}. SoundFont is a widely adopted software-based instrument format for rendering MIDI into audio. Because this rendering is non-differentiable, it cannot pass gradients to VeloEst, which motivates a proxy. Diff-SFProxy is not asked to reproduce waveforms. Instead, it maps note events to loudness-related acoustic parameters: pitch-conditioned harmonic energy (PHE) and onset-window spectral flux (OSF). This design yields two benefits over Diff-Synth: (i) many SoundFonts can plug in off the shelf, whereas Diff-Synth requires a dedicated design per instrument; (ii) gradients focus on velocity-relevant intensity and attack cues rather than waveform details sensitive to recording and timbral mismatch.\footnote{Code is available at \url{https://github.com/zhanh-he/sfproxy-velocity-estimation}}

The contributions of this paper are threefold. First, we formulate label-scarce cross-instrument MIDI velocity estimation and investigate two differentiable adaptation strategies: Diff-Synth and Diff-SFProxy. Second, we introduce Diff-SFProxy itself, a compact SoundFont proxy that supervises on note-wise loudness-related acoustic parameters rather than waveforms. Third, we validate the proposed framework on piano and guitar recordings. To evaluate without the use of ground-truth velocity labels, we adopt Bark-scale specific loudness (BSSL) and Bark-scale total loudness (BSTL)~\cite{zwicker1999psychoacoustics, he2026dynest} to assess whether predicted velocities reproduce the perceived loudness of the original recording, reflecting its performance dynamics.

\section{Cross-Instrument MIDI Velocity Estimation Framework}
\label{sec:framework}
 
\subsection{VeloEst: Pretrained on Piano}
\label{sec:veloest}
Figure~\ref{fig1} shows the overall framework. We adopt the score-HPT model of~\cite{he2026score}, which achieved state-of-the-art performance in MIDI velocity estimation, together with its publicly available checkpoint trained on MAESTRO using ground-truth velocity supervision $\mathcal{L}_{\mathrm{velo}}$. Serving as the shared front-end, VeloEst takes log-Mel spectrograms computed at $100$\,fps from $22050$\,Hz mono audio (FFT size $2048$, hop $221$ samples, Hann window) and aligned note information to estimate MIDI velocities. During cross-instrument adaptation, only VeloEst is updated, while the Diff-Synth and Diff-SFProxy back-ends remain frozen.
 
\subsection{VeloEst Adaptation via Diff-Synth}
\label{sec:adapt_diffsynth}
During the adaptation, only VeloEst parameters are updated. Diff-Synth is a frozen, instrument-specific differentiable synthesizer. Given note events $E$ (pitch, onset, duration) derived from the aligned score and predicted velocities $\hat{V}$, Diff-Synth renders audio $\tilde{a}$. The adaptation loss is the multi-scale spectral loss proposed in \cite{engel2020ddsp}, which is widely adopted in the differentiable synthesizer family \cite{renault2023ddsp_piano, jonason2024guitar, wu2022mididdsp}. We apply it between the rendered audio $\tilde{a}$ and the real performance audio $a$,
\begin{multline}
\label{eq:audio_loss}
\mathcal{L}_{\mathrm{audio}}
= \sum_{\omega \in \Omega} \Bigl(
\bigl\lVert |\mathrm{STFT}_{\omega}(\tilde{a})| - |\mathrm{STFT}_{\omega}(a)| \bigr\rVert_1 \\
+ \bigl\lVert \log(|\mathrm{STFT}_\omega(\tilde{a})|+\epsilon) - \log(|\mathrm{STFT}_\omega(a)|+\epsilon) \bigr\rVert_1
\Bigr),
\end{multline}
where $\Omega=\{2048,1024,512,256,128,64\}$ as in \cite{engel2020ddsp}. For each 
$\omega\in\Omega$, $\mathrm{STFT}_{\omega}$ uses an $\omega$-sample Hann window 
with $75\%$ overlap and an $\omega$-point FFT. We apply $\log(|\cdot|+\epsilon)$ 
with $\epsilon=10^{-7}$ for numerical stability.
 
\subsection{VeloEst Adaptation via Diff-SFProxy}
\label{sec:adapt_diffsfproxy}
As in the Diff-Synth path, only VeloEst parameters are updated. Diff-SFProxy is a frozen differentiable proxy $P$ of an instrument-specific SoundFont. Let $F$ be the note-wise loudness-related acoustic parameter extractor, detailed in Section~\ref{sec:acoustic_params}, that reads audio conditioned on the note events $E$. Given predicted velocities $\hat{v}$, the two feature streams are
\begin{equation}
\label{eq:sfproxy_features}
    Z = F(a, E), \qquad
    \tilde{Z} = P(E, \hat{v}).
\end{equation}
The adaptation loss is the Huber (Smooth-$L_1$) distance between the proxy prediction $\tilde{Z}$ and the extracted loudness-related acoustic parameters $Z$,
\begin{equation}
\label{eq:param_loss}
    \mathcal{L}_{\mathrm{params}}
    = \mathrm{Huber}_\beta\!\left(\tilde{Z} - Z\right),
\end{equation}
with the default $\beta = 1.0$. With $Z_n=(\mathrm{PHE}_n,\mathrm{OSF}_n)$ defined in Section~\ref{sec:acoustic_params}, the Huber loss averages over both components and all valid notes, weighting PHE and OSF equally ($1{:}1$). The Huber form tolerates outliers from onset contamination, such as overlapping notes, while the parameter-space objective concentrates gradients on note intensity and attack behaviour rather than on waveform details weakly related to velocity.

\subsection{Shared Regularizers and Training Objective}
\label{sec:training_objective}
Diff-Synth and Diff-SFProxy adaptation share two auxiliary regularizers on the predicted velocity vector $\hat{v}=(\hat{v}_1,\ldots,\hat{v}_N)\in[0,1]^N$, where $N$ is the number of notes in the segment. The first is an anti-collapse regularizer,
\begin{equation}
\label{eq:anti_loss}
    \mathcal{L}_{\mathrm{anti}}
    = (\bar{v} - \mu)^2
    + \max\!\left(0,\ \sigma_{\min}^2 - \mathrm{Var}(\hat{v})\right),
\end{equation}
where $\bar{v}$ and $\mathrm{Var}(\hat{v})$ are the empirical mean and variance of the predicted velocities, $\mu = 0.5$ is the desired operating point, and $\sigma_{\min}^2 = 0.01$ is the minimum-variance threshold. The mean term resists drift towards extremes, while the hinge variance term penalises collapse towards a single output value. The second is a soft saturation penalty,
\begin{equation}
\label{eq:satu_loss}
    \mathcal{L}_{\mathrm{satu}}
    = \frac{1}{N}\sum_{n=1}^{N}
    \bigl[\max(0,\, \hat{v}_n - \tau)\bigr]^2,
\end{equation}
which quadratically penalises predictions above a near-maximum threshold $\tau = 0.95$, discouraging VeloEst from pushing predictions into the saturated region where audio-derived gradients become uninformative.
\begin{equation}
\label{eq:total_vel_loss}
    \mathcal{L}
    = \mathcal{L}_{\mathrm{main}}
    + \lambda_{\mathrm{anti}}\,\mathcal{L}_{\mathrm{anti}}
    + \lambda_{\mathrm{satu}}\,\mathcal{L}_{\mathrm{satu}},
\end{equation}
where $\mathcal{L}_{\mathrm{main}} \in \{\mathcal{L}_{\mathrm{audio}},\,\mathcal{L}_{\mathrm{params}}\}$ selects the active setting. For both adaptation branches, the loss weights for $\mathcal{L}_{\mathrm{main}}$, $\mathcal{L}_{\mathrm{anti}}$, and $\mathcal{L}_{\mathrm{satu}}$ are fixed at $1{:}0.2{:}0.4$, based on grid search.

\section{Differentiable SoundFont Proxy}
\label{sec:diffproxy}

\subsection{Loudness-Related Acoustic Parameters}
\label{sec:acoustic_params}

To characterize the velocity-dependent acoustic behavior of a SoundFont, we use two note-wise loudness-related parameters adapted from established MIR feature families. The extractor $F$ computes these parameters from the magnitude STFT
$X(k,t)=|\mathrm{STFT}(k,t)|$, using the same FFT settings as Section~\ref{sec:veloest}. Here $k$ indexes frequency bins and $t$ indexes STFT frames. For each valid note $n\in\{1,\ldots,N\}$, the extractor stacks PHE and OSF into
\begin{equation}
\label{eq:Z_def}
Z_n = \bigl(\mathrm{PHE}_n,\ \mathrm{OSF}_n\bigr) \in \mathbb{R}^2,
\end{equation}
and stacks all note-wise pairs into $Z = (Z_1,\ldots,Z_N) \in \mathbb{R}^{N\times 2}$. Each parameter is defined below.
 
\textbf{Pitch-conditioned harmonic energy (PHE)} draws on harmonic-structure-based note-intensity estimation~\cite{ewert2011estimating}. It accumulates magnitude energy around the retained harmonics of the note pitch:
\begin{equation}
\label{eq:harmonic_energy}
\resizebox{\columnwidth}{!}{$
\displaystyle
\mathrm{PHE}_n =
\log\!\left(1 + \sum_{h=1}^{H_n}
\frac{1}{|\mathcal{B}_{n,h}|\,|\mathcal{T}_n|}
\sum_{k \in \mathcal{B}_{n,h}} \sum_{t \in \mathcal{T}_n} X(k,t)^2 \right)
$}
\end{equation}
where $h$ indexes harmonics, $\mathcal{B}_{n,h}$ is the $\pm1$-bin frequency band around the $h$-th harmonic of note $n$, $\mathcal{T}_n$ is the set of STFT frames within $120$\,ms after the note onset, and $H_n\le5$ is the number of retained harmonics below Nyquist. The narrow band handles STFT leakage while rejecting nearby pitches. The five-harmonic cap keeps PHE on lower partials that are less exposed to inharmonicity, tuning or rendering mismatch, and overlap.
 
\textbf{Onset-window spectral flux (OSF)} is based on spectral-flux onset functions, which are used in onset detection by capturing positive spectral differences~\cite{dixon2006onset,boeck2013maximum}. It measures the broadband, half-wave-rectified spectral increase around the note onset:
\begin{equation}
\label{eq:onset_flux}
\resizebox{\columnwidth}{!}{$
\displaystyle
\mathrm{OSF}_n =
\log\!\left(1 + \frac{1}{|\mathcal{O}_n|}
\sum_{t \in \mathcal{O}_n} \sum_k \bigl[ X(k,t) - X(k,t-1) \bigr]_+ \right)
$}
\end{equation}
where $\mathcal{O}_n$ is the set of STFT frames from $20$\,ms before to $80$\,ms after the note onset, and $[u]_+ = \max(u,0)$ for any scalar $u$. The short pre-onset side absorbs alignment slack and provides a local reference. The longer post-onset side covers the main struck or plucked-string transient without folding in sustained energy.
 
\subsection{Proxy Architecture}
\label{sec:proxy_arch}
 
\begin{figure}[htbp]
\centering
\includegraphics[width=\columnwidth]{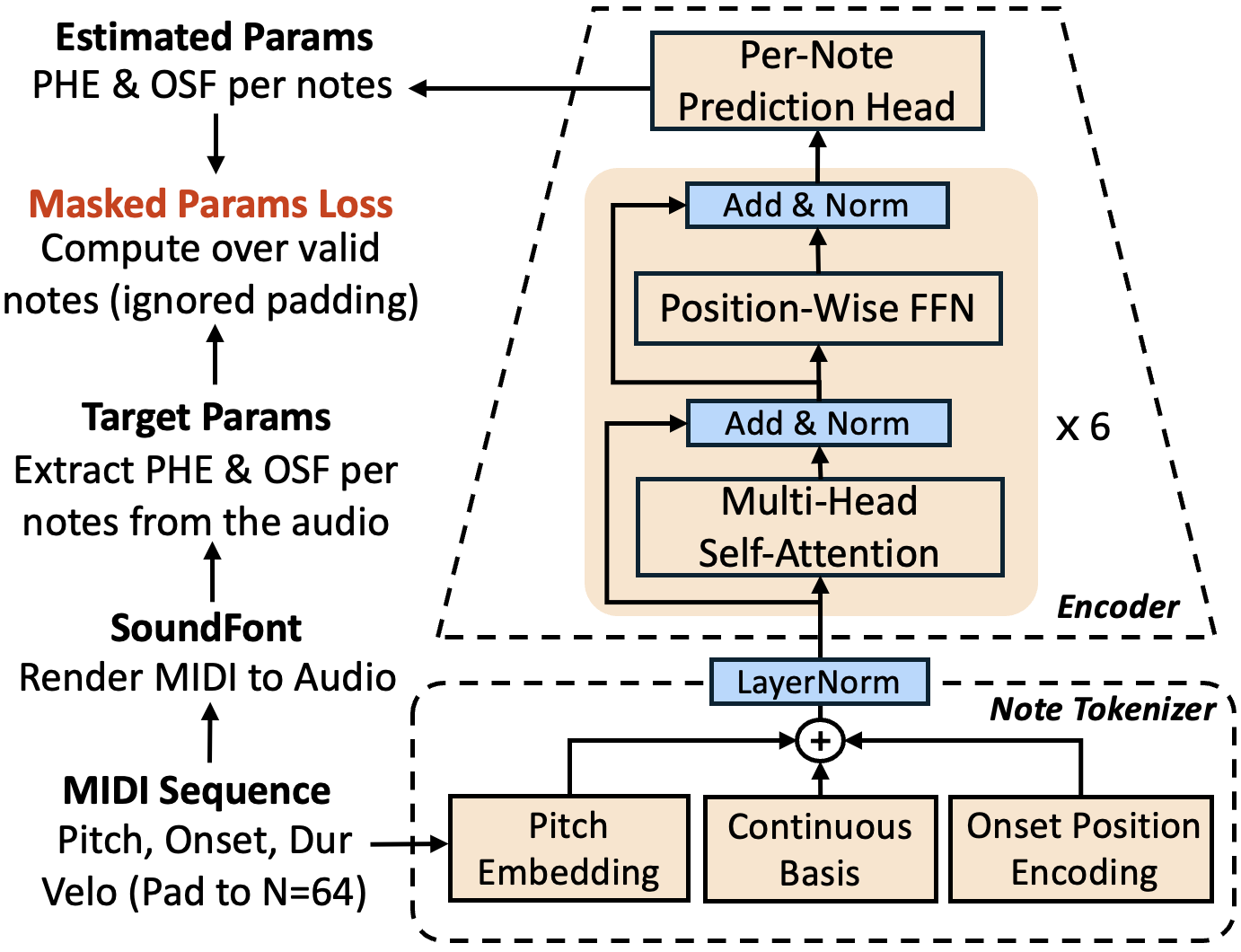}
\caption{An encoder-only Transformer maps padded note sequences to per-note PHE/OSF targets extracted from SoundFont-rendered audio.}
\label{fig:proxy_arch}
\end{figure}

Figure~\ref{fig:proxy_arch} shows the proxy architecture. We use a Transformer encoder that maps a padded sequence of up to $N_{\max}=64$ note events in a $5$\,s segment to the corresponding $(\widetilde{\mathrm{PHE}}_n, \widetilde{\mathrm{OSF}}_n)$ pairs. We choose $N_{\max}=64$ as our data analysis on MAESTRO reveals that human performances rarely exceed 64 notes per 5 seconds. Each note token combines a learnable pitch embedding, a shared learnable basis that maps onset, duration, and velocity (each normalized to $[0,1]$) to additive directions in token space, and a two-layer MLP with GELU activation on the normalized onset acting as a learnable positional encoding. These components are summed and passed through LayerNorm to form the note token, and then padded slots are zeroed. The encoder stacks six pre-norm Transformer layers ($d=256$, $8$ heads, FFN width $4d = 1024$, GELU, dropout $0.1$), with a key-padding mask excluding padded slots from attention. A shared note-wise head ($d\!\to\!d\!\to\!2$, with GELU) emits each $(\widetilde{\mathrm{PHE}}_n, \widetilde{\mathrm{OSF}}_n)$ pair. The encoder-only design fits here because this is a set-to-set regression task.

\subsection{Training Data and Objective}
\label{sec:proxy_data_train}
 
Training the proxy involves three steps: MIDI segments are drawn from samplers, rendered through a SoundFont and extracted to PHE/OSF targets as teacher data, and a loss aligns the proxy's predictions with those targets.
 
\textbf{Note event sampling.}
Note events come from a three-part mixture sampler: (1) $50\%$ of the segments draw notes uniformly across pitch, duration, and inter-onset interval, giving broad coverage of ordinary polyphonic texture; (2) $30\%$ are simple monophonic or low-polyphony passes that isolate how velocity alone affects a single note; (3) $20\%$ is deliberately adversarial, with dense overlapping onsets and large chords, so that the proxy is also exposed to its worst-case inputs during training. Within any chord, per-note velocities are sampled independently so that the proxy learns per-note rather than chord-level responses.
 
\textbf{Teacher data from SoundFonts.}
The teacher audio for the proxy is rendered using a specified SoundFont (\texttt{.sfz}) via the Python package \texttt{sfizz} at $22050$\,Hz. Audio is rendered in segments of $2$, $5$, or $10$\,s with reverb disabled, and short fade-outs are applied to suppress synthesizer state leakage. Rendered segments and their PHE/OSF targets are cached once and reused across epochs, decoupling proxy training from online rendering.
 
\textbf{Training objective.}
Given the sampled note sequence $(E,V)$, the proxy predicts a $(\widetilde{\mathrm{PHE}}_n, \widetilde{\mathrm{OSF}}_n)$ pair for each note. We optimize the network by minimizing the Huber loss in Eq.~\eqref{eq:param_loss}, applying a strict mask to exclude padded sequence slots from the gradient computation.

\section{Experiments}
\label{sec:experiments}
 
\subsection{Datasets and SoundFont Selection}
\label{sec:datasets}
We use piano as the source domain, where velocity labels are available, and guitar as the main label-scarce target domain. For piano, we train on the MAESTRO V3 dataset~\cite{hawthorne2019maestro} ($198.7$\,h) using its default train/valid/test split. For guitar, we adapt on the GAPS dataset~\cite{riley2024gaps} ($14$\,h) using its default train/valid/test split. Furthermore, we evaluate the cross-dataset generalization on the SMD dataset~\cite{smd} ($4.7$\,h) for piano and the Fran\c{c}ois Leduc (FL) dataset~\cite{riley2024high} ($4.0$\,h) for guitar. MAESTRO and SMD consist of Yamaha Disklavier performances. GAPS features nylon-string classical guitar, while FL comprises solo jazz guitar recordings on a mix of nylon-string, steel-string, electric, and acoustic guitars.
 
The SoundFont is chosen deliberately for each target instrument, since a mismatched timbre would produce a training signal that the target recordings cannot match. We audition candidates from the public FreePats sound banks and select the closest match: Salamander Grand Piano~\cite{sf_salamander} for piano, which approximates the Yamaha grand piano timbre of MAESTRO and SMD, and Spanish Classical Guitar~\cite{sf_spanishguitar} for guitar, a close match for GAPS and the nylon-string performance in FL. We also include YDP Grand Piano~\cite{sf_ydpgrand}, an earlier and less realistic Yamaha grand piano sample, to investigate the metric sensitivity to SoundFont change (Section~\ref{sec:discussion}).
 
\subsection{Diff-SFProxy: Velocity-Recovery Diagnostic}
\label{sec:proxy_exp}

\textbf{Experimental rationale.}
Before deploying Diff-SFProxy in cross-instrument adaptation, we test whether its gradients can be trusted. The question is: \emph{do the proxy's gradients point velocity in the right direction?} We answer it with a controlled inversion on synthetic audio, where the true velocities are known. This isolates the gradient pathway from any real-audio domain gap. A pass here is a precondition for Section~\ref{sec:velocity_exp}.

\subsubsection{Training Protocol}
Each proxy is trained on $20$k MIDI segments, with an additional $2$k for validation. The optimizer is AdamW with a learning rate of $10^{-4}$, weight decay $0.01$, batch size $64$, and a fixed random seed $19$. We train Diff-SFProxy for $200$ epochs, and the validation loss typically plateaus between epochs $150$ and $200$. We select the checkpoint with the lowest validation loss for each run. To accommodate the three input segment lengths $2/5/10$\,s, we set $N_{\max}$ to $32/64/128$ accordingly. All three proxies use the same Transformer architecture, resulting in $4.9$\,M parameters each.
 
\subsubsection{Diagnostic Protocol}
We render MIDI segments with the SoundFont and extract their PHE/OSF as the target. We then reset all note velocities to $64$, freeze the proxy, and run gradient descent (Adam, $120$ steps, learning rate $0.03$) on the candidate velocities until the proxy's predicted PHE/OSF match the target. The headline metric is the per-note velocity MAE on the scale $[0, 127]$ between recovered and true velocities. A substantial reduction from the initial error indicates that the proxy's gradients yield update directions toward the true velocities, which is exactly the property needed for adaptation. We use two test groups, each containing $2$k MIDI segments. The \textbf{in-domain test} draws from the same sampler used for training. The \textbf{stress test} uses denser polyphony, larger chords, and tighter inter-onset intervals to probe behavior beyond the training distribution.

\begin{table}[b]
\centering
\small
\renewcommand{\arraystretch}{1.2}
\setlength{\tabcolsep}{5pt}
\begin{tabular}{l||ccc|ccc}
\toprule
& \multicolumn{3}{c}{\textbf{In-domain}} & \multicolumn{3}{c}{\textbf{Stress}} \\
\cmidrule(lr){2-4}\cmidrule(lr){5-7}
\textbf{Diff-SFProxy\,} & Init & Rec. & Gain $\uparrow$ & Init & Rec. & Gain $\uparrow$ \\
\midrule
Piano,  $2$\,s  & 37.2 & 12.3 & 66.9\% & 38.6 & 19.5 & 49.5\% \\
\textbf{Piano,  $5$\,s}  & 37.9 & 12.2 & \textbf{67.8\%} & 38.5 & 17.7 & \textbf{54.0\%} \\
Piano,  $10$\,s & 38.2 & 12.5 & 67.3\% & 38.5 & 17.8 & 53.7\% \\
\midrule
Guitar, $2$\,s  & 33.6 & 9.3 & 72.3\% & 34.2 & 11.3 & 67.0\% \\
\textbf{Guitar, $5$\,s}  & 34.6 &  9.1 & \textbf{73.7\%} & 33.8 & 10.7 & \textbf{68.3\%} \\
Guitar, $10$\,s & 34.6 &  9.5 & 72.5\% & 33.8 & 10.8 & 68.0\% \\
\bottomrule
\end{tabular}
\caption{Velocity recovery diagnostic. Init and Rec.\ report per-note velocity MAE in MIDI units $[0, 127]$ before and after optimization; Gain $= (\text{Init}-\text{Rec.})/\text{Init}\times 100\%$.}
\label{tab:proxy_recovery}
\end{table}

\subsubsection{Results}
Table~\ref{tab:proxy_recovery} reports the diagnostic for piano and guitar proxies at $2/5/10$\,s segment lengths. In-domain, all six proxies cut the initial error to about a third. Under the stress test, recovery improves from $2$\,s to $5$\,s, most clearly on piano. From $5$\,s to $10$\,s the gain disappears, with all cells within half a MIDI unit. We therefore use the $5$\,s proxy as the default in the cross-instrument adaptation, and keep $2$\,s as an ablation. The $10$\,s setting adds training cost without further gain, so we discard it.

\begin{table*}[htbp]
    \centering
    \setlength{\tabcolsep}{2.6pt}
    \small
    \renewcommand{\arraystretch}{1.2}
    \begin{tabular}{l||ccc|ccc||cc|cc}
        \toprule
        \multicolumn{1}{c||}{\multirow{2}{*}{\textbf{Method}}}
        & \multicolumn{6}{c||}{\textbf{Piano}}
        & \multicolumn{4}{c}{\textbf{Guitar}} \\
        \cmidrule(lr){2-7} \cmidrule(l){8-11}
        & \multicolumn{3}{c|}{\textbf{MAESTRO test set}}
        & \multicolumn{3}{c||}{\textbf{SMD dataset}}
        & \multicolumn{2}{c|}{\textbf{GAPS test set}}
        & \multicolumn{2}{c}{\textbf{FL dataset}} \\
        \cmidrule(lr){2-4} \cmidrule(lr){5-7} \cmidrule(lr){8-9} \cmidrule(l){10-11}
        & MAE\textsubscript{Velo} $\downarrow$
        & $r$\textsubscript{BSSL} $\uparrow$
        & $r$\textsubscript{BSTL} $\uparrow$
        & MAE\textsubscript{Velo} $\downarrow$
        & $r$\textsubscript{BSSL} $\uparrow$
        & $r$\textsubscript{BSTL} $\uparrow$
        & $r$\textsubscript{BSSL} $\uparrow$
        & $r$\textsubscript{BSTL} $\uparrow$
        & $r$\textsubscript{BSSL} $\uparrow$
        & $r$\textsubscript{BSTL} $\uparrow$ \\
        \midrule
        \textbf{Flat Velocity:} set at 64 & 14.8 & 0.757 & 0.622 & 15.4 & 0.701 & 0.582 & 0.737 & 0.701 & 0.714 & 0.587 \\
        \textbf{VeloEst:} pretrained on MAESTRO
        & \textbf{3.9} & \textbf{0.878} & \textbf{0.892}
        & \textbf{8.0} & \textbf{0.829} & \textbf{0.816}
        & 0.769 & 0.798 & 0.740 & 0.699 \\
        
        \textbf{VeloEst + Diff-Synth} & 19.2 & 0.776 & 0.752 & 19.7 & 0.789 & 0.742 & 0.669 & 0.624 & 0.646 & 0.547 \\
        \,-\, uses 2s backend instead of 5s
        & 25.0 & 0.702 & 0.553 & 25.9 & 0.662 & 0.520 & 0.760 & 0.770 & 0.729 & 0.636 \\
        \midrule
        \textbf{VeloEst + Diff-SFProxy} (proposed) \,\, & \underline{10.5} & \underline{0.869} & \underline{0.881}
        & \underline{11.3} & \underline{0.821} & \underline{0.806}
        & \textbf{0.794} & \textbf{0.860} & \textbf{0.777} & \textbf{0.788} \\
        \,-\, uses 2s backend instead of 5s
        & 10.9 & 0.858 & 0.870 & 11.8 & 0.820 & 0.804
        & \underline{0.787} & \underline{0.836} & \underline{0.771} & \underline{0.756} \\
        \,-\, w/o anti-collapse and saturation loss
        & 12.9 & 0.789 & 0.698 & 14.4 & 0.735 & 0.653 & 0.740 & 0.705 & 0.723 & 0.614
        \\
        \,-\, w/o VeloEst pretrained weights
        & 20.5 & 0.770 & 0.757 & 21.2 & 0.723 & 0.703 & 0.702 & 0.686 & 0.683 & 0.600 \\
        \bottomrule
    \end{tabular}%
    \caption{Evaluation on piano and guitar datasets. Diff-Synth and Diff-SFProxy backends process 5 s segments by default. Best and second-best results are marked in bold and underlined, respectively.}
    \label{tab:unified_results}
\end{table*}
 
\subsection{Cross-Instrument VeloEst Adaptation}
\label{sec:velocity_exp}

\textbf{Experimental rationale.}
Our experiment on adaptation has two purposes. On guitar, where velocity labels are unavailable, it provides a convenient label-free approach to MIDI velocity estimation.
On piano, where ground-truth velocity is available on MAESTRO and SMD, it serves as a controlled diagnostic: we adapt on MAESTRO so that predictions can be checked against true labels, verifying whether our non-velocity supervision pushes velocity toward the right value. The piano result is therefore not a claim about improving over a supervised piano baseline. It is a sanity check that the supervision signal is informative.

\subsubsection{Training Protocol} 
We compare three settings on the same VeloEst front-end: the piano-pretrained baseline (no adaptation), Diff-Synth adaptation, and Diff-SFProxy adaptation. The Diff-Synth backends \cite{renault2023ddsp_piano, jonason2024guitar} were originally implemented for $3$\,s input segments. We retrain each at $2$\,s and $5$\,s with their original hyperparameters, then freeze them.
 
During adaptation, VeloEst processes $10$\,s segments, while the frozen backend operates on $5$\,s crops by default and $2$\,s for ablation. Backend losses are averaged over valid notes in each crop and backpropagated to the corresponding VeloEst $10$\,s predictions, keeping the gradient scale comparable across backend crop lengths. Adaptation is run separately on the MAESTRO and GAPS training sets for $120$k iterations using Adam with learning rate $10^{-4}$, batch size $8$, and step decay $0.9$ every $10$k iterations. SMD and FL are used only for evaluation. All runs use seed $19$, and the final checkpoint is selected by validation $r_{\mathrm{BSSL}}$ on MAESTRO for piano or GAPS for guitar.

\subsubsection{Evaluation Metrics}
For piano datasets with ground-truth velocities, we follow \cite{kim2024method, he2026score} and report per-note $\mathrm{MAE}_{\mathrm{Velo}}$. For datasets without velocity labels, $\mathrm{MAE}_{\mathrm{Velo}}$ cannot be computed. We instead re-render the aligned notes through the selected SoundFont using the predicted velocities, and compare the resynthesis with the original recording in perceptual loudness space.

We use Bark-scale specific loudness (BSSL) and Bark-scale total loudness (BSTL)~\cite{zwicker1999psychoacoustics, he2026dynest} as perceptually grounded comparison features. BSSL is a frame-wise critical-band representation, while BSTL is its band-summed contour, tracking perceived loudness over time. Both features capture the main perceptual effect targeted by velocity estimation, namely the temporal shaping of loudness, while reducing sensitivity to timbral and recording-condition mismatch between real audio and SoundFont resynthesis. We avoid two natural alternatives. Raw RMS energy ignores frequency-dependent perceptual sensitivity. Note-wise loudness measures, including PHE and OSF, are onset-locked, miss the continuous loudness contour, and coincide with the proxy training targets, making the evaluation circular.

Pearson correlation ($r$) between real and rendered loudness contours is our primary metric. Because it is invariant to global scale and offset, it reduces sensitivity to loudness differences introduced by gain, microphone response, or renderer calibration. We denote Pearson correlation computed on BSSL and BSTL as $r_{\mathrm{BSSL}}$ and $r_{\mathrm{BSTL}}$, respectively. Cosine similarity ($cs$) and MAE are retained as diagnostic alternatives: cosine similarity is scale-invariant but not offset-invariant, whereas MAE is sensitive to both. Their behaviour is examined in Section~\ref{sec:discussion}.

\subsubsection{Results}
Table~\ref{tab:unified_results} reports evaluation across all four test sets. The row labeled `VeloEst: pretrained on MAESTRO' is the supervised checkpoint applied without further training, serving as the zero-shot transfer baseline for SMD, GAPS, and FL. On piano, this checkpoint remains the strongest supervised reference. Both adaptation strategies underperform on piano because they optimize indirect performance-matching objectives rather than the available velocity labels.

On guitar, Diff-SFProxy achieves the best $r_{\mathrm{BSSL}}$ and $r_{\mathrm{BSTL}}$ on both GAPS and FL, outperforming flat velocity, zero-shot transfer, and Diff-Synth under both backend crop lengths. Diff-Synth does not reliably improve guitar performance and can degrade it: on FL, the $5$\,s Diff-Synth backend falls below flat velocity in $r_{\mathrm{BSSL}}$ ($0.646$ vs.\ $0.714$). This supports the concern in Section~\ref{sec:adapt_diffsynth}: waveform-level objectives are more exposed to timbral and room-acoustic mismatch than parameter-level supervision. The crop-length effect also differs sharply between the two approaches. On FL, increasing the backend crop from $2$\,s to $5$\,s reduces Diff-Synth's $r_{\mathrm{BSSL}}$ by $0.083$, whereas Diff-SFProxy changes by only $0.006$. Longer crops expose waveform-level objectives to more residual mismatch, while parameter-level supervision remains focused on note-wise velocity-related cues. The zero-shot piano checkpoint already exceeds flat velocity on both guitar sets, suggesting partial cross-instrument transfer, while Diff-SFProxy provides a consistent further improvement.
 
\subsubsection{Ablations}
The bottom rows of Table~\ref{tab:unified_results} isolate two design choices. Removing $\mathcal{L}_{\mathrm{anti}}+\mathcal{L}_{\mathrm{satu}}$ gives a plausible $\mathrm{MAE}_{\mathrm{Velo}}$ on MAESTRO, but $r_{\mathrm{BSSL}}$ falls near the flat-velocity baseline. The outputs collapse toward a near-constant value, consistent with the median-seeking behavior of MAE: central predictions can keep absolute error tolerable while losing the dynamic contour. Removing the piano-pretrained checkpoint leaves the proxy loss without a supervised velocity anchor, causing performance to drift across all datasets and underperform flat velocity on three of the four guitar metrics. Piano pretraining therefore acts as a crucial prior for cross-instrument adaptation.

\section{Discussion}
\label{sec:discussion}

\textbf{Scope and metric implications.}
Velocity is not the only factor that shapes rendered audio loudness: pedaling, continuous controllers, and effects such as reverb also contribute \cite{ryu2024midfild}. Diff-Synth backends such as DDSP-Piano expose several of these controls. Consequently, when adaptation updates velocity alone while the remaining controls are fixed or unobserved, the velocity gradient can be confounded by residuals attributable to those controls. Diff-SFProxy instead restricts the optimization to velocity, making adaptation more tractable. This simplification comes at a cost: the recovered velocity may absorb effects that, in the real performance, arose from other controls. Hence, the recovered velocity is not a canonical property of the acoustic performance. It is defined relative to the chosen SoundFont as the velocity that best reproduces the perceptual dynamics of the original audio. This renderer-conditioned velocity is therefore the relevant target for label-scarce expressive analysis or resynthesis. Accordingly, the evaluation metric must be robust to SoundFont choice and sensitive to dynamics differences.

\textbf{Why Pearson $r$ on BSSL/BSTL.}
Table~\ref{tab:metric_diagnostic} tests metrics against the two requirements implied above: sensitivity to expressive dynamics and robustness to renderer-specific loudness calibration. To test dynamics sensitivity, $\Delta(\mathrm{GT}-\mathrm{Flat\,64})$ asks whether a metric can tell expressive resynthesis from a flat one. Among the similarity metrics, Pearson $r$ gives the larger separation on both BSSL ($+0.118$) and BSTL ($+0.263$). Cosine similarity also orders GT above flat, but severely compresses the margin ($+0.052$ on BSSL and $+0.047$ on BSTL), proving \textbf{insensitive} to fine dynamics differences. To evaluate renderer robustness, $\Delta\lvert\mathrm{Sal.}-\mathrm{YDP}\rvert$ asks whether a metric is stable when the rendering SoundFont is swapped. Here, MAE proves \textbf{over-sensitive} to absolute gain: it drifts by $0.193$ on BSSL and $1.723$ on BSTL, substantially more than the corresponding drift on $r$ ($0.034$, $0.011$). On real recordings, this drift would conflate dynamics differences with renderer choice. Overall, Pearson $r$ on BSSL/BSTL provides the ideal trade-off, whereas cosine similarity and MAE fail on sensitivity and robustness, respectively, and are kept only as diagnostic alternatives.

\begin{table}[htbp]
\centering
\setlength{\tabcolsep}{0.7pt}
\small
\renewcommand{\arraystretch}{1.2}
\begin{tabular}{l||ccc|ccc}
    \toprule
    \multicolumn{1}{c||}{\multirow{3}{*}{\textbf{Method}}}
    & \multicolumn{6}{c}{\textbf{MAESTRO test set}} \\
    \cmidrule(l){2-7}
    & \multicolumn{3}{c|}{\textbf{BSSL}} & \multicolumn{3}{c}{\textbf{BSTL}} \\
    \cmidrule(lr){2-4} \cmidrule(l){5-7}
    & $r$ & $cs$ & MAE & $r$ & $cs$ & MAE \\
    \midrule
    \textbf{GT velocity}            & 0.870 & 0.939 & 0.702 & 0.883 & 0.975 & 3.706 \\
    \,-\, uses YDP Piano            & 0.836 & 0.926 & 0.510 & 0.872 & 0.973 & 1.983 \\
    $\Delta\lvert$Sal.\,$-$\,YDP$\rvert$ $\downarrow$
                                    & \underline{0.034} & \textbf{0.013} & 0.193
                                    & \underline{0.011} & \textbf{0.002} & 1.723 \\
    \midrule
    \textbf{Flat velocity = 64}     & 0.752 & 0.887 & 0.786 & 0.620 & 0.928 & 4.065 \\
    \,-\, uses YDP Piano \,            & 0.733 & 0.883 & 0.637 & 0.617 & 0.930 & 2.750 \\
    $\Delta\lvert$Sal.\,$-$\,YDP$\rvert$ $\downarrow$
                                    & \underline{0.019} & \textbf{0.004} & 0.149
                                    & \underline{0.003} & \textbf{0.002} & 1.314 \\
    \midrule
    \textbf{$\Delta$ (GT $-$ Flat) $\uparrow$}
                                    & \textbf{+0.118} & \underline{+0.052} & -0.083
                                    & \textbf{+0.263} & \underline{+0.047} & -0.359 \\
    \bottomrule
\end{tabular}
\caption{MAESTRO metric diagnostic. Real audio is compared with GT-note resynthesis using GT or flat velocity under the Salamander (default) or YDP Grand Piano. Summary rows show GT--Flat separation and SoundFont drift.}
\label{tab:metric_diagnostic}
\end{table}

\section{Conclusion}
In this paper, we presented a framework for cross-instrument MIDI velocity estimation. A piano-pretrained estimator is adapted to target instruments without velocity labels through one of two differentiable interfaces: waveform-space Diff-Synth or loudness-related acoustic parameter-space Diff-SFProxy. The proposed Diff-SFProxy leverages note-wise loudness parameters extracted from black-box SoundFont renders, focusing the gradient on velocity-dependent behavior rather than waveform details. On guitar, where velocity labels are unavailable, Diff-SFProxy improves over zero-shot transfer and waveform-level adaptation, while Diff-Synth often degrades performance. Our finding suggests that, when the task is to estimate a single control parameter rather than reconstruct audio, parameter-wise supervision through a renderer proxy is more reliable than waveform supervision.

Future work includes extending the framework to other instruments, such as violin and wind instruments. Automating SoundFont selection through timbre matching would replace the current by-ear procedure and scale the method to broader datasets. Beyond velocity, the framework could also recover other expressive control parameters, such as sustain pedaling in piano and bow-pressure curves in violin.

\bibliography{ISMIR2026}

@string{aaai = "Proc. AAAI"}

@string{aaai = "Proceedings of the AAAI Conference on Artificial Intelligence (AAAI)"}

@string{icassp = "Proc. ICASSP"}

@string{icassp = "Proc. IEEE Int. Conf. Acoust., Speech, Signal Process. (ICASSP)"}

@string{icassp = "Proceedings of the IEEE International Conference on Acoustics, Speech and Signal Processing (ICASSP)"}

@string{ismir = "Proc. ISMIR"}

@string{ismir = "Proc. Int. Soc. Music Inf. Retr. Conf. (ISMIR)"}

@string{ismir = "Proceedings of the International Society for Music Information Retrieval Conference (ISMIR)"}

@string{smc = "Proc. SMC"}

@string{smc   = "Proc. Sound Music Comput. Conf. (SMC)"}

@string{smc = "Proceedings of the Sound and Music Computing Conference (SMC)"}

@string{dafx = "Proc. DAFx"}

@string{dafx = "Proceedings of the International Conference on Digital Audio Effects (DAFx)"}

@string{icmc = "Proc. ICMC"}

@string{icmc = "Proceedings of the International Computer Music Conference (ICMC)"}

@string{mmsp = "Proc. MMSP"}

@string{mmsp = "Proceedings of the IEEE International Workshop on Multimedia Signal Processing (MMSP)"}

@string{icml = "Proc. ICML"}

@string{icml = "Proceedings of the International Conference on Machine Learning (ICML)"}

@string{iclr = "Proc. ICLR"}

@string{iclr = "Proceedings of the International Conference on Learning Representations (ICLR)"}

@string{waspaa = "Proc. WASPAA"}

@string{waspaa = "Proceedings of the IEEE Workshop on Applications of Signal Processing to Audio and Acoustics (WASPAA)"}

@string{jaes = "J. Audio Eng. Soc."}

@string{jaes = "Journal of the Audio Engineering Society"}

@string{appsci = "Appl. Sci."}

@string{appsci = "Applied Sciences"}

@InProceedings{smd,
  author={M. M{\"u}ller and V. Konz and W. Bogler and V. Arifi-M{\"u}ller},
  title={Saarland Music Data (SMD)},
  booktitle={Late-Breaking and Demo Session of the 12th Int. Conf. on Music Information Retrieval},
  year={2011},
  bk_organization={ISMIR},
}

@InProceedings{hawthorne2019maestro,
  author={Hawthorne,Curtis and Stasyuk,Andriy and Roberts,Adam and Simon,Ian and Huang,Cheng-Zhi Anna and Dieleman,Sander and Elsen,Erich and Engel,Jesse and Eck,Douglas},
  title={Enabling Factorized Piano Music Modeling and Generation with the MAESTRO Dataset},
  booktitle=iclr,
  year={2019},
  mycomment={MAESTRO数据集的文章, 记得去overleaf上确认下正确引用格式
        URL: https://openreview.net/forum?id=r1lYRjC9F7},
}

@InProceedings{ryu2024midfild,
  author={Ryu, Jesung and Rhyu, Seungyeon and Yoon, Hong-Gyu and Kim, Eunchong and Yang, Ju Young and Kim, Taehyun},
  title={MID-FiLD: MIDI Dataset for Fine-Level Dynamics},
  booktitle=aaai,
  year={2024},
  pages={222--230},
  bk_address={Vancouver, Canada},
  mycomment={Article available: https://ojs.aaai.org/index.php/AAAI/article/view/27774
        Github available: https://github.com/POZAlabs/MID-FiLD_code
        Demo available: https://pozalabs.github.io/MID-FiLD_demo},
}

@InProceedings{riley2024gaps,
  author={Riley, Xavier and Guo, Zixun and Edwards, Andrew C. and Dixon, Simon},
  title={{GAPS:} A Large and Diverse Classical Guitar Dataset and Benchmark Transcription Model},
  booktitle=ismir,
  year={2024},
  pages={611--617},
  bk_address={San Francisco, California, USA},
  mycomment={Article available: https://arxiv.org/abs/2408.08653
        Github available: https://github.com/aim-qmul/GAPS
        Demo available: https://aim-qmul.github.io/GAPS/},
}

@InProceedings{morsi2025enable_rach3,
  author={Morsi, Alia and Chiruthapudi, Suhit and Peter, Silvan and Pilkov, Ivan and Bishop, Laura and Maezawa, Akira and Serra, Xavier and Cancino-Chacón, Carlos Eduardo},
  title={Enabling Empirical Analysis of Piano Performance Rehearsal with the Rach3 MIDI Dataset},
  booktitle=ismir,
  year={2025},
  pages={484--491},
  bk_address={Daejeon, South Korea},
  mycomment={rach3的进一步扩展,
             Article available: https://repositori.upf.edu/items/7d341d5f-d12e-464a-93c4-06143cece675
             Github unavailable.},
}

@InProceedings{riley2024high,
  author={Riley, Xavier and Edwards, Drew and Dixon, Simon},
  title={High Resolution Guitar Transcription via Domain Adaptation},
  booktitle=icassp,
  year={2024},
  pages={1051--1055},
  mycomment={Francois Leduc Dataset URL: https://zenodo.org/records/10984521},
}

@InProceedings{tamer2023violin,
  author={Tamer, Nazif Can and {\"O}zer, Yigitcan and M{\"u}ller, Meinard and Serra, Xavier},
  title={High-Resolution Violin Transcription Using Weak Labels},
  booktitle=ismir,
  year={2023},
  pages={223--230},
  bk_address={Milan, Italy},
  mycomment={Article available: https://archives.ismir.net/ismir2023/paper/000025.pdf
        Github available: https://github.com/MTG/violin-transcription/},
}

@InProceedings{he2026dynest,
  author={He, Zhanhong and Meng, Hanyu and Huang, Defeng and Togneri, Roberto},
  title={Joint Estimation of Piano Dynamics and Metrical Structure with a Multi-task Multi-Scale Network},
  booktitle=icassp,
  year={2026},
  pages={14607--14611},
}

@InProceedings{kim2024method,
  author={Kim, H. and Serra, X.},
  title={A method for MIDI velocity estimation for piano performance by a {U}-Net with attention and {FiLM}},
  booktitle=ismir,
  year={2024},
  bk_month={Nov.},
  pages={304--310},
  bk_address={San Francisco, USA},
  bk_organization={ISMIR},
  mycomment={URL: https://repositori.upf.edu/handle/10230/61103},
}

@InProceedings{he2026score,
  author={He, Zhanhong and Togneri, Roberto and Huang, Defeng},
  title={Score-Informed Transformer for Refining MIDI Velocity in Automatic Music Transcription},
  booktitle=smc,
  year={2026},
  note={Accepted, to appear},
  mycomment={Article available: https://arxiv.org/abs/2508.07757
        Github available: https://github.com/zhanh-he/score-informed-amt-velocity
        Demo not available.},
}

@inproceedings{borovik2025symupe,
  author    = {Borovik, Ilya and Gavrilev, Dmitrii and Viro, Vladimir},
  title     = {{SyMuPe}: Affective and Controllable Symbolic Music Performance},
  booktitle = {Proceedings of the 33rd ACM International Conference on Multimedia},
  series    = {MM '25},
  pages     = {10699--10708},
  year      = {2025},
  address   = {Dublin, Ireland},
  publisher = {Association for Computing Machinery},
  doi       = {10.1145/3746027.3755871},
  mycomment={Article available: https://doi.org/10.1145/3746027.3755871
        Github available: https://github.com/ilya16/SyMuPe
        Demo available: https://ilya16.github.io/SyMuPe/},
}

@InProceedings{tang2025integrated,
  author={Tang, Jingjing and Cooper, Erica and Wang, Xin and Yamagishi, Junichi and Fazekas, Gy{\"o}rgy},
  title={Towards An Integrated Approach for Expressive Piano Performance Synthesis from Music Scores},
  booktitle=icassp,
  year={2025},
  bk_pages={1--5},
  doi={10.1109/ICASSP49660.2025.10890623},
  mycomment={Github available: https://github.com/tangjjbetsy/S2A
        Demo available: https://tangjjbetsy.github.io/S2A/},
}

@InProceedings{simonetta2022acoustics,
  author={Simonetta, Federico and Ntalampiras, Stavros and Avanzini, Federico},
  title={Acoustics-specific Piano Velocity Estimation},
  booktitle=mmsp,
  year={2022},
  mycomment={Article available: https://arxiv.org/abs/2203.16294
        Github available: https://github.com/LIMUNIMI/ContextAwareAMT
        Demo not available.
        Article available: https://arxiv.org/abs/2203.16294
        Github available: https://github.com/LIMUNIMI/ContextAwareAMT
        Demo not available.},
}

@inproceedings{zhang2022atepp,
  author    = {Zhang, Huan and Tang, Jingjing and Rafee, Syed Rifat Mahmud and Dixon, Simon and Fazekas, Gy{\"o}rgy},
  title     = {{ATEPP}: A Dataset of Automatically Transcribed Expressive Piano Performance},
  booktitle = ismir,
  year      = {2022},
  pages     = {446--453},
}

@InProceedings{rhyu2022sketching,
  author={Rhyu, S. and Kim, S. and Lee, K.},
  title={Sketching the Expression: Flexible Rendering of Expressive Piano Performance with Self-Supervised Learning},
  booktitle=ismir,
  year={2022},
  pages={178--185},
  bk_address={Bengaluru, India},
  bk_organization={ISMIR},
  mycomment={DOI: https://doi.org/10.5281/zenodo.7342916
        URL: https://doi.org/10.5281/zenodo.7342916},
}

@InProceedings{jeong2018timbre,
  author={Jeong, Dasaem and Kwon, Taegyun and Nam, Juhan},
  title={A Timbre-Based Approach to Estimate Key Velocity From Polyphonic Piano Recordings},
  booktitle=ismir,
  year={2018},
  pages={120--127},
  bk_address={Paris, France},
  bk_organization={ISMIR},
  mycomment={URL: https://archives.ismir.net/ismir2018/paper/000196.pdf},
}

@Article{zhang2024dexter,
  author={Zhang, Huan and Chowdhury, Shreyan and Cancino-Chac{\'o}n, Carlos Eduardo and Liang, Jinhua and Dixon, Simon and Widmer, Gerhard},
  title={Dexter: Learning and controlling performance expression with diffusion models},
  journal=appsci,
  year={2024},
  volume={14},
  number={15},
  pages={6543},
  publisher={MDPI},
}

@InProceedings{dan2006psy,
  author={Dannenberg, Roger B.},
  title={The Interpretation of MIDI Velocity},
  booktitle=icmc,
  year={2006},
  pages={193--196},
  mycomment={URL: https://www.cs.cmu.edu/~rbd/papers/velocity-icmc2006.pdf},
}

@Article{combes2025neural,
  author={Combes, Paolo and Weinzierl, Stefan and Obermayer, Klaus},
  title={Neural Proxies for Sound Synthesizers: Learning Perceptually Informed Preset Representations},
  journal=jaes,
  year={2025},
  bk_month={September},
  volume={73},
  number={9},
  pages={561--577},
  mycomment={Article available: https://arxiv.org/abs/2509.07635
        Github available: https://github.com/pcmbs/synth-proxy
        Demo not available.},
}

@Article{renault2023ddsp_piano,
  author={Renault, Lenny and Mignot, R{\'e}mi and Roebel, Axel},
  title={{DDSP-Piano}: A Neural Sound Synthesizer Informed by Instrument Knowledge},
  journal=jaes,
  year={2023},
  volume={71},
  number={9},
  pages={552--565},
  mycomment={URL: https://aes.org/journal-online/?num=9&vol=71},
}

@InProceedings{jonason2024guitar,
  author={Jonason, Nicolas and Wang, Xin and Cooper, Erica and Juvela, Lauri and Sturm, Bob L. T. and Yamagishi, Junichi},
  title={{DDSP-Based} Neural Waveform Synthesis of Polyphonic Guitar Performance From String-Wise MIDI Input},
  booktitle=dafx,
  year={2024},
  pages={208--215},
  bk_address={Guildford, United Kingdom},
  mycomment={Article available: https://www.dafx.de/paper-archive/2024/papers/DAFx24_paper_49.pdf
        Github available: https://github.com/erl-j/ddsp-guitar
        Demo available: https://erl-j.github.io/neural-guitar-web-supplement/},
}

@InProceedings{barkan2023is2,
  author={Barkan, Oren and Shvartzman, Shlomi and Uzrad, Noy and Laufer, Moshe and Elharar, Almog and Koenigstein, Noam},
  title={{InverSynth II}: Sound Matching via Self-Supervised Synthesizer-Proxy and Inference-Time Finetuning},
  booktitle=ismir,
  year={2023},
  pages={642--648},
  bk_address={Milan, Italy},
  mycomment={Article available: https://archives.ismir.net/ismir2023/paper/000076.pdf},
}

@InProceedings{martinez2021differentiable,
  author={Mart\'inez Ram\'irez, Marco A. and Wang, Oliver and Smaragdis, Paris and Bryan, Nicholas J.},
  title={Differentiable Signal Processing With Black-Box Audio Effects},
  booktitle=icassp,
  year={2021},
  pagess={66-70},
  bk_month={June},
  mycomment={Article available: https://arxiv.org/abs/2105.04752
        Github available: https://github.com/adobe-research/DeepAFx
        Demo not available.},
}

@InProceedings{engel2020ddsp,
  author={Engel, Jesse and Hantrakul, Lamtharn and Gu, Chenjie and Roberts, Adam},
  title={{DDSP:} Differentiable Digital Signal Processing},
  booktitle=iclr,
  year={2020},
  mycomment={Article available: https://arxiv.org/abs/2001.04643
        Github available: https://github.com/magenta/ddsp
        Demo available: https://goo.gl/magenta/ddsp-demo},
}

@Book{zwicker1999psychoacoustics,
  author={Zwicker, Eberhard and Fastl, Hugo},
  title={Psychoacoustics: Facts and Models},
  year={1999},
  volume={22},
  address={Berlin},
  publisher={Springer},
  series={Springer Series in Information Sciences},
  edition={2nd, updated},
  mycomment={DOI: https://doi.org/10.1007/978-3-662-11962-1},
}

@misc{sf_salamander,
  author       = {Holm, Alexander},
  title        = {Acoustic Grand Piano: {Salamander Grand Piano}},
  howpublished = {FreePats Project sound bank},
  year         = {2020},
  month        = jun,
  note         = {Version V3+2020-06-02; Creative Commons Attribution 3.0; accessed 2026-04-18},
  url          = {https://freepats.zenvoid.org/Piano/acoustic-grand-piano.html}
}

@misc{sf_ydpgrand,
  author       = {{FreePats Project}},
  title        = {Acoustic Grand Piano: {{YDP} Grand Piano}},
  howpublished = {FreePats Project sound bank},
  year         = {2016},
  month        = aug,
  note         = {Version 2016-08-04; Creative Commons Attribution 3.0; accessed 2026-04-18},
  url          = {https://freepats.zenvoid.org/Piano/acoustic-grand-piano.html}
}

@misc{sf_spanishguitar,
  author       = {{FreePats Project}},
  title        = {Nylon-String Acoustic Guitar: {Spanish Classical Guitar}},
  howpublished = {FreePats Project sound bank},
  year         = {2019},
  month        = jun,
  note         = {Version 2019-06-18; Creative Commons CC0 1.0 public domain dedication; accessed 2026-04-18},
  url          = {https://freepats.zenvoid.org/Guitar/acoustic-guitar.html}
}

@InProceedings{wu2022mididdsp,
  author={Wu, Yusong and Manilow, Ethan and Deng, Yi and Swavely, Rigel and Kastner, Kyle and Cooijmans, Tim and Courville, Aaron and Huang, Cheng-Zhi Anna and Engel, Jesse},
  title={{MIDI-DDSP}: Detailed Control of Musical Performance via Hierarchical Modeling},
  booktitle=iclr,
  year={2022},
  mycomment={Article available: https://openreview.net/pdf?id=UseMOjWENv
        Github available: https://github.com/magenta/midi-ddsp
        Demo available: https://midi-ddsp.github.io/},
}

@InProceedings{wang2026vioptt,
  author={Wang, Ting-Kang and Peng, Yueh-Po and Su, Li and Cheung, Vincent K. M.},
  title={VioPTT: Violin Technique-Aware Transcription from Synthetic Data Augmentation},
  booktitle=icassp,
  year={2026},
  pages={15947--15951},
  bk_address={Barcelona, Spain},
  mycomment={Article available: https://arxiv.org/abs/2509.23759
        Github available: https://github.com/y10ab1/VioPTT
        Demo available: https://y10ab1.github.io/VioPTT/},
}

@inproceedings{ewert2011estimating,
  author    = {Ewert, Sebastian and M{\"u}ller, Meinard},
  title     = {Estimating Note Intensities in Music Recordings},
  booktitle = icassp,
  year      = {2011},
  pages     = {385--388}
}

@inproceedings{dixon2006onset,
  author    = {Dixon, Simon},
  title     = {Onset Detection Revisited},
  booktitle = dafx,
  year      = {2006},
  pages     = {133--137},
}

@inproceedings{boeck2013maximum,
  author    = {B{\"o}ck, Sebastian and Widmer, Gerhard},
  title     = {Maximum Filter Vibrato Suppression for Onset Detection},
  booktitle = dafx,
  year      = {2013},
  pages ={55--61},
}

@inproceedings{sato2024annotationfree,
  author    = {Sato, Gakusei and Akama, Taketo},
  title     = {Annotation-Free Automatic Music Transcription with Scalable Synthetic Data and Adversarial Domain Confusion},
  booktitle = {Proceedings of the 2024 IEEE International Conference on Multimedia and Expo ({ICME})},
  pages     = {1--6},
  year      = {2024},
  doi       = {10.1109/ICME57554.2024.10688348}
}

@InProceedings{maman2022unaligned,
  title     = {Unaligned Supervision for Automatic Music Transcription in The Wild},
  author    = {Maman, Ben and Bermano, Amit H.},
  booktitle = icml,
  pages     = {14918--14934},
  year      = {2022},
}

@inproceedings{zang2024synthtab,
  title     = {{SynthTab}: Leveraging Synthesized Data for Guitar Tablature Transcription},
  author    = {Zang, Yongyi and Zhong, Yi and Cwitkowitz, Frank and Duan, Zhiyao},
  booktitle = icassp,
  pages     = {1286--1290},
  year      = {2024},
}

@inproceedings{kusaka2025learn,
  title     = {Learn from Virtual Guitar: A Comparative Analysis of Automatic Guitar Transcription Using Synthetic and Real Audio},
  author    = {Kusaka, Yuta and Maezawa, Akira},
  booktitle = waspaa,
  pages     = {1--5},
  year      = {2025},
}
 
\end{document}